\documentstyle[11pt]{article}
\input{epsf}

\newcommand\pTwo[3]{
  \begin{figure}[ht]\leavevmode
  \epsfxsize=8cm\epsfysize=8cm\epsfbox{#1.ps}
  \epsfxsize=8cm\epsfysize=8cm\hfill\epsfbox{#2.ps}
  \vspace{-3cm}\caption{\small #3}\label{#1}
\end{figure}  }

\def\beq{\begin{eqnarray}}    
\def\eeq{\end{eqnarray}}      

\def\tr{\,\mbox{tr}\,}                  
\def\Tr{\,\mbox{Tr}\,}                  

\def\al{\alpha}
\def\be{\beta}

\def\ga{\gamma}
\def\de{\delta}
\def\fr{\frac}

\def\ze{\zeta}

\def\la{\lambda}
\def\na{\nabla}
\def\pa{\partial}

\def\si{\sigma}

\def\ph{\varphi}

\def\De{\Delta}
\def\La{\Lambda}

\begin{document}

\begin{center}
{\LARGE\sl
Back reaction of vacuum and the renormalization group flow}
\vskip 3mm
{\LARGE\sl
from the conformal fixed point}
\vskip 8mm

{\bf G. Cognola}
\footnote{e-mail address: cognola@science.unitn.it}
\vskip 1mm\vskip 1mm
{\sl Dipartimento di Fisica, Universit\`a di Trento, 38050, Povo (Trento) and}

\vskip 1mm
{\sl
Istituto Nazionale di Fisica Nucleare, Gruppo Collegato di Trento, Italia}
\vskip 1mm\vskip 1mm\vskip 1mm\vskip 1mm

{\bf I. L. Shapiro}
\footnote{e-mail address:
shapiro@ibitipoca.fisica.ufjf.br}
\vskip 1mm\vskip 1mm
{\sl Departamento de Fisica -- ICE,
Universidade Federal de Juiz de Fora\\
36036-330, Juiz de Fora -- MG - Brazil}
\vskip 1mm
{\sl  and Tomsk State Pedagogical University, Tomsk, 634041, Russia }
\vskip 1mm\vskip 1mm\vskip 1mm\vskip 1mm

\vskip 8mm
\end{center}

\bigskip
\hfill April 1998
\bigskip

\noindent
{\Large \it Abstract.}
We consider the GUT-like  model with two scalar fields
which has infinitesimal deviation
from the conformal invariant fixed point at high energy region.
In this case the dominating quantum effect is the conformal trace
anomaly and the interaction between the anomaly-generated propagating
conformal factor of the metric and the usual dimensional scalar
field. This interaction leads to the renormalization group flow
from the conformal point.
In the supersymmetric conformal invariant model such an effect
produces a very weak violation of sypersymmetry at lower energies.


\setcounter{page}1
\renewcommand{\thefootnote}{\arabic{footnote}}
\setcounter{footnote}0
\vskip 8mm

\section{Introduction}

The formulation of consistent quantum field theory in curved space-time
requires two new elements: the action of vacuum $\,S_v[g_{\mu\nu}]\,$
and also the nonminimal scalar-curvature interaction
$\,\frac12\xi R\ph^2\,$
for each scalar field. The theory which lacks one of those elements
is nonrenormalizable, because the corresponding counterterms appear
anyway, and already at the one-loop level (see, for example,
\cite{book} for the introduction). One can suppose that such a theory
with an action $S = S_{min} + S_{non-min} + S_v$ should be valid until
the effects of quantum gravity become important, that is for the
energies well below the Planck scale. It is naturally to suppose that
at the very high energies below the Planck scale the theory doesn't
depend on the massive parameter and is conformal invariant. The condition
of local conformal invariance is the special choice of the nonminimal
parameter $\,\xi=\frac16\,$. The necessary action of vacuum has,
in the conformal case, the restricted form:
\beq
S_v = \int d^4x \sqrt{-g}\,(\, a_1C^2+a_2E+a_3\Box R \,)\,,
\label{vac}
\eeq
where $\,\,\,\sqrt{-g}\,C^2 \,=\,
\sqrt{-g}\,C_{\mu\nu\al\be}\,C^{\mu\nu\al\be}\,\,\,$
is the conformal
invariant square of the Weyl tensor and

\noindent
$E=R^2_{\mu\nu\al\be}-4R^2_{\mu\nu}+R^2\,\,$ is the Gauss-Bonnet integrand.
On quantum level the conformal invariance is spoiled by the trace
anomaly which takes place in both vacuum \cite{duff} and
matter-fields \cite{bracol} sectors of the theory. In the vacuum sector
the trace
anomaly arises (in the framework of dimensional regularization and
the minimal subtraction scheme) because the counterterms, which
must be added to the one-loop effective action, are local expressions
in $n$-dimensional space-time. In general, the origin
of the anomaly is the lack of regularization preserving both
general covariance and local conformal invariance (see \cite{duff-rev}
for the review of conformal anomaly and related issues).

The interest to the conformal invariant theories in four dimensions ($4d$)
is partially
inspired by the important role which they play in the two-dimensional
($2d$) theories which are traditionally considered as a toy models for
the realistic higher dimensional theories. There are, however, a serious
differences between $2d$ and $4d$ theories. First of all,
in $2d$ the fields with a different spin contribute to the anomaly
(which comes from the $\,\,\int d^2x\sqrt{-g}R\,$-type counterterm)
with a different sign, that leads to the existence of the
critical dimension for the
sigma-models and thus in $2d$ the anomaly may be canceled.
In $4d$ there are three possible types of the one-loop counterterms
(\ref{vac}) and the fields with the  different spin
give contributions with the same sign to the renormalization constants
of the vacuum parameters $a_1$ and $a_2$ in (\ref{vac}).
Therefore in $4d$ the critical dimension doesn't exist.
The situation remains the same, even if the supersymmetry and quantum
effects of gravity use to be incorporated \cite{frts-consugra}.
Another possibility which may be realized only in $2d$ is to start
from the non-linear sigma-model in a background fields (defined as
a geometric objects in target space) and thus provide
the cancelation of anomaly in the noncritical dimension
\cite{ftstr,cfmp}. In $4d$ this
scheme doesn't work either, because the nonlinear sigma-model is
non-renormalizable and its possible higher derivative version
\cite{buke} presumably contains unphysical massive ghosts.
Thus, the quantum effects inevitably lead to the violation of conformal
invariance through the trace anomaly with the consequent propagation
of the conformal factor \cite{rei,frts}\footnote{The derivation of the
anomaly-generated effective action of the conformal factor has been also
performed in \cite{buodsh} for the case of the theory with torsion
and in \cite{buku} for the supersymmetric matter on the background
of simple supergravity.}. The study of the quantum theory
of the conformal factor as a low-energy version of quantum gravity
has been started in \cite{antmot} (See also \cite{first}).

In the matter fields sector of the classically conformal invariant
theory the one-loop divergences are also invariant but the finite
one-loop contributions to the one-loop effective action are not.
This can be,
in particular, seen through the renormalization of the composite operators
in the expression for the energy-momentum tensor \cite{bracol}.
This renormalization leads
to the violation of the conformal invariance in the higher-loop
divergences \cite{hath}. Thus, in $4d$, the classically conformal
invariant theory suffers from two kinds of deceases: the conformal
anomaly arises at one loop and produces the propagation of the conformal
factor and also the nonconformal divergences take place at higher loops
and break the renormalizability. Therefore the local conformal symmetry
in $4d$ can not be exact, and can be realized only as an approximate
high energy phenomena which may be called asymptotic conformal invariance.
The asymptotic conformal invariance has been originally discovered
in \cite{buod,buch} as a consequence of the conformal invariance
of the one-loop divergences. This kind of the
asymptotic conformal invariance doesn't follow from the asymptotic
freedom and it puts some extra constraints on the
multiplet composition and on the values of the coupling constants of the
gauge model \cite{bushya}. The shortcoming of this approach is that
the higher order corrections to the $\be$-functions spoil the asymptotic
conformal invariance and hence $\,\xi = \frac16\,$ is not the
renormalization group fixed point beyond the one-loop level.
Here we adopt another point of view
on the asymptotic conformal invariance. Let us suppose that the
asymptotically free or finite gauge theory in curved space-time, which
is generated at Planck energy scale as an effective low-energy theory, is
originally conformal invariant and thus
the conformal invariant theory is the initial condition for the
renormalization group flow in far UV rather than its fixed point.

For the theories with a weak coupling the one-loop effects are always
dominating and hence the leading quantum effect
is the conformal (trace) anomaly
$$
\,<T_{\mu}^{\;\;\mu}>=k_1C^2+k_2E+k_3\Box R.\,
$$
The values of $k_1,_2,_3$ depend only on the
number of fields of different spin in a GUT model \cite{duff}.
The anomaly leads to the equation for the effective action
$\,2 g_{\mu\nu}\,\delta W /{\delta g_{\mu\nu}} =
-\sqrt{-g}\,T^\mu_\mu .\,\,$
The solution of this equation is the $4d$ analog of the Polyakov action.
It can be written in a local form \cite{rei} as:
\beq
W[g_{\mu\nu},\sigma] = S_{c}[g_{\mu\nu}]+
\int d^4x\sqrt{-g}\,\left\{\,\fr12\sigma\Delta\sigma
+\si\, [k'_1C^2+k'_2\left(E-\fr23\Box R\right)\,]
- \frac{1}{12}\,k'_3R^2 \,\right\}
\label{3}.
\eeq
The values of the coefficients $\,k'_{1,2,3}\,$ are related to the ones
of the trace anomaly in a well known way \cite{rei}.
The propagator of the conformal factor $\,\si\,$ is an inverse to the
fourth derivative conformal invariant operator \cite{pan,rei,frts}:
\beq
\Delta=\Box^{2}+2R^{\mu\nu}\nabla_{\mu}\nabla_{\nu}-\fr23R\Box
+\fr13(\nabla^{\mu}R)\nabla_{\mu}
\label{4}
\eeq
and hence in the flat space limit $\si$ is massless field which
can be essential at long distances.
The solution (\ref{3}) includes also an arbitrary conformal invariant
functional $S_{c}$. In four dimensional quantum field theory
this functional can not be calculated exactly, and one can only
establish its high energy asymptotic or derive some lower order
terms in its expansion. However this functional doesn't
depend on the conformal factor and thus it is irrelevant for our
purposes
\footnote{ The action of gravity induced by conformal anomaly
has interest for the construction of quantum gravity \cite{first,deser1}.
It looks possible that the proper choice of the functional
$S_c$ may revoke the discrepancy between the above effective action
and direct calculations of $W[g_{\mu\nu},\sigma]$ in \cite{vilk} and
with the results of the study of the two- and three-point functions
\cite{osborn,deser}.}.

$\,$
>From the physical point of view anomaly means the propagating conformal
factor. Simultaneously the two-loop effects produce the violation of
the conformal constraint $\,\xi=\frac16\,$ and as a result the
conformal factor starts to interact with the usual scalar field.
The investigation of the physical consequences of this interaction
has been started in \cite{foo}.
The interaction with the propagating
conformal factor modifies the renormalization group equation for
the scalar coupling $f$ and $\xi$. These equations have minimal
IR-stable fixed point $\xi=f=0$, and in the vicinity of this point
one meets the first order phase transition,
as a result the Einstein gravity is induced \cite{foo}. In \cite{foo}
the theory with one scalar field was considered. Here we
are going to perform the detailed study of the renormalization group
equations in the framework of more complicated models with two scalar
multiplets. Our purpose is to estimate the running
of the couplings due to the interaction with the anomaly-generated
conformal factor and, in particular, explore the renormalization group
flow from the conformal fixed point in the models which are finite
and supersymmetric in flat space time.

Starting from the conformal initial point in far UV limit one can
trace the renormalization group flows for the coupling constants and
$\,\xi\,$ backward to the lower energies.
In this way one can predict the value of $\,\xi\,$ at the
lower energy scales.
Thus we arrive at the necessity to study the renormalization group
behavior in the theory which has only infinitesimal deviation from
the conformal fixed point in far UV. To perform this
in a consistent way one has to take into account the effect
of conformal anomaly and the consequent quantum effects of the
propagating conformal factor \cite{foo}. The detailed study of the
renormalization group behavior is the purpose of the present paper.

The paper is organized in a following way. In the next section we
formulate the model of GUT coupled to the propagating conformal factor.
In section 3 the derivation of the one-loop counterterms is performed.
Section 4 is devoted to the renormalization group
equations for two different models, and in the last section we draw
our conclusions.

\section{Interaction of matter fields with conformal factor}

We suppose that the effects of conformal factor
are relevant below the Planck scale where the nonminimal parameters
$\xi$ become slightly different from $1/6$ due to the higher loop
effects. In curved space-time the transition to the low energies (or
long distances) corresponds to conformal transformation in the induced
gravity action
\cite{book}, after that classical fields and induced gravity appear in
a different conformal points \cite{foo} (see also \cite{antmot}).
In order
to take this into account one has to make a conformal transformation
of the metric in (\ref{3}) and then consider the unified theory.
However it is more useful to perform the conformal transformation
in the action of the matter
fields. Such a transformation corresponds to some change of variables
in the path integral for the unified theory, and doesn't modify the
results of quantum calculations in the matter field sector, which
we are interested in.

The only source of conformal noninvariance in the action of the
massless GUT model is the nonminimal term in the scalar sector.
As far as the values of the parameters
$\xi$ are not equal to $\frac{1}{6}\,$ the
conformal factor starts to interact with the scalar fields.
Consider the general $\,SU$(N)$\,$ model
with two types of scalar
fields: real ones $\Phi^a$ in the adjoint representation and
complex ones $\varphi^i$ in vector representation of the gauge group.
In curved space-time one has to introduce two nonminimal parameters
$\xi_1$ and $\xi_2$ -- one for each type of the scalar fields.
This model has been investigated in \cite{shya} without taking into
account the back reaction of vacuum, and one can find more complete
information including the full set
of the $\beta$-functions in this paper. Below we write down
only that parts of the $\be$-functions which we shall actually use.

Introducing the scale parameter $\alpha$ we arrive at the following
action for the conformal factor coupled to scalar fields:
$$
S_{sc}\,=\,W[g_{\mu\nu},\sigma]+\int d^{4}x \sqrt{-g}\,
\left\{ \left[\fr12\,(1-6\xi_1)\Phi^a\Phi^a +
(1-6\xi_2)\ph_i^+ \ph^i\right]
\left(\alpha^2 (\na\si)^2 +\alpha\Box\sigma\right)
\right.
$$
\beq
\left.
+ \fr12\,g^{\mu\nu}\,({\cal D}_\mu\Phi)^a\,({\cal D}_\nu\Phi)^a
+ g^{\mu\nu}\,({\cal D}_\mu \ph^+) _i \, ({\cal D}_\nu\ph)^i
+ \fr12\,\xi_1\,R\,\Phi^a\Phi^a + \xi_2\,R\,\ph_i^+ \ph^i
- V(\Phi^a, \ph^i)\right\},
\label{action}
\eeq
where $W[g_{\mu\nu},\sigma]$ has been defined in (\ref{3}) and we use
the notation
$(\na\si)^2 = g^{\mu\nu}\partial_{\mu}\sigma\partial_{\nu}\sigma$.
In all the expressions ${\cal D}$ are the derivatives of the
matter fields which are covariant with respect to both gauge
and gravitational fields.
The notations for the $SU$(N) group including the relations between
symmetric $D_{rab}$ and antisymmetric $f_{abc}$ structure constants,
generators $\,\left(\frac{\la}{2}\right)_i^j\,$, traces etc. can be found
in \cite{book} (see also second reference in \cite{bush}).
The flat-space part of the potential has the form
$$
V(\Phi^a, \ph^i) = \frac18\,f_1\,(\Phi^a\Phi^a)^2
+ \frac18\,f_2\,(\Phi^aD_{rab}\Phi^b)^2
+ \frac12\,f_3\,(\Phi^a\Phi^a)\,(\ph_i^+ \ph^i)
$$
\beq
+ \frac12\,f_4\,(\Phi^aD_{rab}\Phi^b)^2\,\ph_i^+
\left(\frac{\la}{2}\right)^i_j\ph^j
+ \frac12\,f_5\,(\ph_i^+ \ph^i)^2\,.
\label{potential}
\eeq
The action of scalar fields (\ref{action}) must be supplemented by the
action of spinors and gauge fields which are part of the GUT model. The
corresponding Lagrangian has the form \cite{shya}
$$
{\cal {L}} \,=\, -\frac14\,(G_{\mu\nu}^a)^2
+ i\,\sum_{k=1}^{m}{\bar \Psi}^a_{(k)}
\left(\ga^\mu {\cal D}_\mu^{ab} - h_1 f^{abc}\Phi^c\right)\Psi^b_{(k)}
+ i\sum_{k=1}^{m}{\bar \psi}^i_{(k)}
[\ga^\mu {\cal D}_\mu^{ij} - h_2 \left(\frac{\la^a}{2}\right)^{j}_i
\Phi^a]\psi_{j,(k)} \,+
$$
\beq
+\,i\,\sum_{s=1}^{n}{\bar \chi}^i_{(s)}\ga^\mu
{\cal D}_\mu^{ij}\chi^j_{(s)} + ih_3\sum_{k=1}^{m}[\,
{\bar \psi}^i_{(k)}\left(\frac{\la^a}{2}\right)^{i}_j
\ph^{+j}\Psi^a_{(k)}
-{\bar \Psi}^a_{(k)}\left(\frac{\la^a}{2}\right)^{j}_i
\ph^{i}\psi_{j,(k)}]\,.
\label{rest}
\eeq

We are interested in the quantum theory of matter fields
and conformal factor $\si$ on the background of the classical metric.
The interaction between matter fields and conformal factor
arises as a result of the conformal transformation of the metric
$g_{\mu\nu}\to g'_{\mu\nu}=g_{\mu\nu}\exp(2\alpha\sigma)$
and the matter field
$w \rightarrow w' = w\exp(d_{w}\alpha\sigma)$,
where $d_{w}$ is the conformal weight of the field
$w$. One can see that the massless spin-$\frac12$ and spin-1
fields decouple from the conformal factor.
The only kind of fields which takes part in such an interaction are
scalars, for which the interaction with conformal factor appears as a
result of the violation of the condition $\xi=1/6$  at low energies.
The contributions of other matter fields
to the $\beta$-functions and
effective potential of the scalar fields do not depend on the conformal
factor, they are given by the usual expressions \cite{shya,book}.

\section{Calculation of the one-loop divergences}

The calculation of the one-loop divergences of the theory (\ref{action})
is not a trivial problem because the classical action
contains second derivative terms as well
as fourth derivative ones. Here we shall use the approach of Ref.
\cite{bush}, where
the one-loop divergences have been calculated for higher derivative
quantum gravity coupled to the matter fields.
Following \cite{bush}, we shall use the background field method and
generalized Schwinger-DeWitt technique of \cite{bavi}.
Thus we start from the separation of fields into background
$\sigma,\phi$ and quantum $\tau,\eta$ ones by changing
$(\sigma,\Phi^a,\ph^i,\ph^+_i)
\longrightarrow ({\sigma}',{\Phi^a}',{\ph^i}',{\ph_i^+}')$ where
\beq
\sigma' = \sigma + \tau,\,\,\,\,\,\,\,
{\Phi^a}' = \Phi^a + i\eta^a,\,\,\,\,\,\,\,
{\ph^i}' = \ph^i + i\chi^i,\,\,\,\,\,\,\,
{\ph_i^+}' = \ph_i^+ + i\chi_i^+
\label{6}
\eeq
and the imaginary units are introduced for convenience.
The one-loop effective action is defined as
\beq
\Gamma = \frac{i}{2} \Tr\ln {\hat {H}}
\label{7},
\eeq
where ${\hat{H}}$ is the bilinear (with respect to the quantum fields
$\tau,\eta,\chi,\chi^+$) form of the classical action (\ref{action}).
After some algebra we get the following self-adjoint bilinear form
\beq
\hat{H}=\left(\matrix{
H_{\tau\tau} &H_{\tau\eta} &H_{\tau\chi} &H_{\tau\chi^+}   \cr
H_{\eta\tau} &H_{\eta\eta} &H_{\eta\chi} &H_{\eta\chi^+}   \cr
H_{\chi^+\tau} &H_{\chi^+\eta} &H_{\chi^+\chi} &H_{\chi^+\chi^+}   \cr
H_{\chi\tau} &H_{\chi\eta} &H_{\chi\chi} &H_{\chi\chi^+}
\cr} \right)
\label{8}\:,
\eeq
with the following derivative structure of the operator $\hat{H}$:
\beq
\hat{H} =\left(\matrix{
\Box^{2}+
2V^{\mu\nu}\nabla_{\mu}\nabla_{\nu}+ N^{\mu}\nabla_{\mu}+U
&\qquad {\hat Q}_1\Box + {\hat Q}_2^{\mu}\nabla_{\mu} +
{\hat Q}_3\cr
{\hat P}_1\Box + {\hat P}_2^{\mu}\nabla_{\mu} + {\hat P}_3
&\qquad {\hat 1}\Box + {\hat E}^\mu\nabla_\mu+ {\hat D}
}\right)
\label{10}\:.
\eeq
The bilinear form of the action is matrix differential operator
with fourth derivatives
in the $H_{\tau\tau}\,$ sector and with
second derivatives in the sector of usual scalar fields and in the
mixed pieces.
This structure of the bilinear form is similar to the one
which is known from
the theory of multiscalar GUT coupled to higher derivative gravity
\cite{bush}.
This is indeed natural, because what we are doing now is nothing
but the study of induced quantum gravity (\ref{3}) unified
with the same GUT
model. This analogy facilitates the calculations considerably, because
of the following three reasons:
\vskip 1mm
\noindent
i) The general expression for the divergent part of
$\frac{i}{2} \Tr\ln {\hat{H}}\,$(\ref{10})
is known from \cite{bush}
$$
\Gamma_{div}=\frac{\mu^{n-4}}{\varepsilon}\,2
\int d^n x\sqrt{-g}\,\tr\,\left\{
\,\frac14 {\hat P}_2^\mu {\hat Q}_{2\mu}+\fr14 {\hat P}_1 {\hat Q}_3
- \frac14 V^{\mu}_{\mu} {\hat P}_1 {\hat Q}_1
- {\hat D} {\hat P}_1 {\hat Q}_1 + \frac12 {({\hat P}_1 {\hat Q}_1)}^2
\right.
$$$$
\left.
+ \frac12 {\hat Q}_2^\mu\nabla_\mu {\hat P}_1
- \frac16 R {\hat P}_1 {\hat Q}_1
+ \frac1{24} V^{\mu\nu} V_{\mu\nu} + \frac1{48} (V^{\mu}_{\mu})^2
- \frac16 V^{\mu\nu} R_{\mu\nu} + \frac1{12} V^\mu_\mu R - U
\right.
$$
\beq
\left.
+\frac1{20} \left(R^{\mu\nu}R_{\mu\nu}-\frac13R^2\right)
-\frac1{36}R^2 \right\}
+ \frac{i}{2}\Tr\ln\,\left\{{\hat 1} {\Box} + {\hat E}^\la\nabla_\la
+ {\hat D} \right\}_{div}
+\mbox{(surface terms)}
\label{12}\:.
\eeq
The above formula includes the standard contribution of the second order
operator
$$\,\,\frac{i}{2}\Tr\ln\,\left\{{\hat 1} {\Box} +
{\hat E}^\la\nabla_\la + {\hat D} \right\}_{div}\,\,
$$
(see, for example,
\cite{book}), the contribution of the minimal fourth order operator
first calculated in \cite{frts} and also the contributions from the
mixed sector derived in Ref. \cite{bush} for the case of higher
derivative gravity coupled to matter fields.
The formula (\ref{12}) shows, in particular, that we do not need
explicit expression for $\,N^\la$.
\vskip 1mm

\noindent
ii) Calculating the divergences with the use of the formula (\ref{12})
one can learn the form of the $beta$-functions. However, since the
formula (\ref{12}) is the same as for the higher derivative gravity,
it is easy to see that the general structure of the
renormalization group equations in our theory is also the same
as the one established in \cite{bush}. In particular, the quantum gravity
corrections to the beta-functions of the nonminimal parameters $\xi_j$
are completely universal, they do not depend on the model and have the
same form as for the one-scalar model \cite{foo}. Therefore all that we
need is the form of the quantum gravity corrections to the beta-functions
of the constants $f_i$. These corrections can be calculated on flat
background and hence in what follows we put $g_{\mu\nu}=\eta_{\mu\nu}$
and switch off all the curvature dependent terms.

\noindent
iii) The structure of the expression (\ref{12}) is direct generalization
of the one we have already studied in \cite{foo} for the 1-scalar model.
One can easily check that here, just as in the 1-scalar model, the
contributions of $\,\tr {\hat P}_1{\hat Q}_1,\,
\,\tr ({\hat P}_1{\hat Q}_2 +  {\hat P}_2{\hat Q}_1),\,
\tr ({\hat P}_1{\hat Q}_3 +  {\hat P}_3{\hat Q}_1),\,$ and
$\,\tr {\hat P}_2{\hat Q}_2\,$ cancel, and therefore we need only the
expressions for
$ V^{\mu\nu},\,U,\,{\hat P}_1,\,{\hat Q}_1,\,{\hat E}^\la,\,{\hat D}$.
Disregarding the curvature dependent terms one can obtain the
following expressions for $\,V,U,{\hat Q}_1,{\hat P}_1,{\hat D}\,$:
$$
V^{\mu\nu} =
- \al^2\,(\ze_1\,\Phi^a\Phi^a + \ze_2\,\ph^+_i\ph^i)\,g^{\mu\nu}\,,
\,\,\,\,\,\,\,\,\,\,\,\,U=0
\:,
$$
$$
\hat{P}_1 = -i \al\left(\matrix{\ze_1\Phi^a \cr
\ze_2\ph^+_i \cr \ze_2\ph^i \cr
\cr} \right)\:,\,\,\,\,\,\,\,\,\,\,\,\,\,\,\,\,
\hat{Q}_1 = -i \al\left(\matrix{\ze_1\Phi^b &
\ze_2\ph^j & \ze_2\ph^+_j \cr} \right)\:,
$$
\beq
{\hat D}\,=\,
\hat{H}=\left(\matrix{
-\ze_1 Z \de^{ab}
+ V^{ab}&
V^a_j & V^{aj} \cr V^i_b & -\ze_2 Z \de^i_j + V^j_i & V^{ij} \cr
V_{ib} & V_{ij} & \ze_2 Z \de^j_i + V^j_i \cr}
\right),
\eeq
where $\,\,Z = \al^2(\na\si)^2+\al({\Box}\si)\,\,$ and
$$
V_{ab} =
\frac{f_1}{2} \Phi^2\de^{ab} + f_1\Phi^a\Phi^b
+ f_2 \left(
\frac12  D_{rab}D_{rcd} + D_{rac}D_{rbd}\right) \Phi^c \Phi^d +
f_3\de^{ab}\ph^2 +
f_4 D_{rab}\,\ph_i^+\left(\frac{\la^r}{2} \right)^i_j\ph^j,
$$$$
V_{aj} =  f_3\Phi^a\ph^+_j +
f_4 D_{rac} \Phi^c \ph_k^+\left(\frac{\la^r}{2}\right)_j^k
,\,\,\,\,\,\,\,\,\,\,\,\,\,\,\,\,\,\,\,
V_{ib} =  f_3\Phi^b\ph^+_i +
f_4 D_{rbc} \Phi^c \ph_k^+\left(\frac{\la^r}{2}\right)_i^k,
$$$$
V_{a}^j =  f_3\Phi^a\ph^j +
f_4 D_{rac} \Phi^c \left(\frac{\la^r}{2}\right)_k^j \ph^k
,\,\,\,\,\,\,\,\,\,\,\,\,\,\,\,\,\,\,\,
V_{b}^i =  f_3\Phi^a\ph^i +
f_4 D_{rac} \Phi^c \left(\frac{\la^r}{2}\right)_k^i \ph^k,
$$$$
V_{j}^i =\frac12 f_3\Phi^c\Phi^c\de^i_j +
\frac12 f_4 \Phi^c D_{rcd} \Phi^d \left(\frac{\la^r}{2}\right)_j^i
+ f_5 \ph^2\de^i_j + f_5\ph^+_j\ph^i,
$$$$
V^{ij} = f_5\,\ph^i\ph^j,
\,\,\,\,\,\,\,\,\,\,\,\,\,\,\,\,\,\,\,\,\,\,\,\,\,\,\,\,\,\,\,\,\,\,
V_{ij} = f_5\,\ph^+_i\ph^+_j\,.
$$

Substituting these expressions into (\ref{12}) we arrive at the
explicit form of the corrections to the matter sector of
$\Gamma_{div}$ from the quantum conformal factor\footnote{The
renormalization in the vacuum sector was discussed in \cite{foo}
for the case of the one-scalar model.
Its form doesn't depend on the number of scalar fields and that
is why we do not discuss it here.}.
$$
\Gamma_{div}^{(1)}
\,=\,\frac{\mu^{n-4}}{\varepsilon}\,2\,\int d^{n}x\sqrt{-g}
\left\{
\frac{1}{2}\alpha^4 (1-\ze_1)^2\,(\Phi^a\Phi^a)^2
+{2}\alpha^4\ze_2^2 (1 - \ze_2^2)(\ph_i^+\ph^i)^2
+ \frac32\,f_1\,\al^2\ze_1^2(\Phi^a\Phi^a)^2 +
\right.
$$$$
\left.
+2\al^4\ze_1\ze_2 (1 - \ze_1 - \ze_2 + \ze_1\ze_2)\Phi^2\ph^+\ph
+ \frac32f_2\al^2\ze_1^2(\Phi^a D_{abc} \Phi^b)^2
+ f_3\al^2\left[\ze_1^2 + 4\ze_1\ze_2 + \ze_2^2\right]\Phi^2\ph^+\ph
\right.
$$\beq
\left.
+ f_4\,\al^2\left[\ze_1^2 + 4\ze_1\ze_2 + \ze_2^2\right]
(\Phi^a D_{abc} \Phi^b)\,\ph^+_i\left(\frac{\la^c}{2}\right)^i_j\ph^j
\,+ \,6f_5\,\al^2\ze_2^2(\ph^+\ph)^2\right\}\,,
\label{divergency}
\eeq
where $\varepsilon = (4\pi)^2(n - 4)$ is the parameter of dimensional
regularization, and we disregarded all surface and matter-independent
terms.

One can see that there are no any divergences
which lead to renormalization of scalar fields $\Phi,\,\ph$. Thus the
only modifications due to the contributions of the conformal factor are
in the renormalization of couplings $f_{1,2,..,5}$ and parameters
$\xi_{1,2}$, and the renormalization group equations include only
$\beta-$functions, but not $\gamma$-functions. The  contribution
of the conformal factor to the effective potential of the scalar fields
depends on the $\beta$-functions only.

\section{Renormalization group equations}

Now we are in a position to perform the renormalization
group study of the back-reaction of vacuum to the matter fields.
For our purposes it is convenient to use
the formulation of the renormalization group in curved space-time,
given in \cite{buch,book}. The renormalization group
equation for the effective action in curved space-time has the form
\beq
\left\{\, \mu\frac{d}{d\mu}+\beta_{p} \frac{d}{d {p}}
-\int d^4\sqrt{-g}\,\gamma_w\,\frac{\delta}{\delta w(x)}
\,\right\}\,\Gamma[w,p,g_{\mu\nu},\mu]=0\,,
\label{15}\eeq
where $w$ is the full set of the quantum fields (gauge, spinor, Higgs
and conformal factor $\si$) and $p$ -- complete set of parameters
including $\xi$'s. The solution corresponding to the desirable scaling
behavior has the form
\beq
\Gamma[e^{-2t}g_{\mu\nu},\,w,\,p,\,\mu]\, =\,
\Gamma [g_{\mu\nu},\,w(t),\,p(t),\,\mu]
\label{16}\,,
\eeq
where $\mu$ is the dimensional parameter of renormalization.
Effective fields and coupling constants obey the equations
\beq
(4\pi)^{2}\frac{dw(t)}{dt} = (\gamma_{w} + d_w)\phi\:,\,\,
\,\,\,\,\,\,\,\,\,\,\,\,\,\,\,\,\,w(0)=w\,,
\eeq
\beq
(4\pi)^{2}\frac{dp(t)}{dt} = \beta_p + p \,d_p,\,\,\,\,\,\,\,\,\,\,\,\,\,\,\,\,
\,\,\,\,\,\,\,\,\,\,\,\,\,\,\,\,p(0) = p\,,
\label{17}
\eeq
where $\gamma$ and $\beta$
functions are defined as usual. According to the results of the previous
section the renormalization of the fields and all couplings except
$f_i$ and $\xi_j$ are not modified by the contributions of the quantum
field $\sigma$. Thus the $\be$-functions for the gauge and
Yukawa coupling constants are just the
ones derived in \cite{shya}. On the other hand, these contributions to the
$\,\be_{f_i}$ and $\,\be_{\xi_j}$ all have universal form and do not
depend on the gauge group of the theory. Thus we find for our theory
(from this moment we will be
using the variables $\zeta_j = 1 -6 \xi_j$ for convenience):
\beq
\be_{f_i} = \be_{f_i}^{(0)} + \De\be_{f_i}
\,,\,\,\,\,\,\,\,\,\,\,\,\,
\be_{\ze_j} = \be_{\ze_j}^{(0)} + \De\be_{\ze_j}
\,,\,\,\,\,\,\,\,\,\,\,\,\,
\be_{h_k} = \be_{h_k}^{(0)}
\,,\,\,\,\,\,\,\,\,\,\,\,\,
\be_{g} = \be_{g}^{(0)},
\label{structure}
\eeq
where $\be_{p}^{(0)}$ is the $\be$-function for the effective parameter
$p$ in curved space-time without
back reaction of vacuum (or other form of quantum gravity) and
$\De\be_{p}$ are the quantum gravity corrections. In our
case, contrary to the high derivative gravity \cite{bush}
those $\De\be_{p}$'s are nonzero only for the scalar and the
nonminimal parameters.

According to recent communications (see, for example, \cite{runinfla})
the values of $\xi_j$ are very important at the energies
between the Planck scale $M_{Pl}=10^{19}GeV$ and the unification
scale $M_{X}=10^{14}GeV$, because there is a hope to meet natural
inflation for this rate of character energies.
Thus our purpose is to evaluate the running of $\xi_j$ (or $\zeta_j$)
backward from the high energy Planck scale. As a lower end of the
energy interval one can take the unification point $M_{X}=10^{14}GeV$
but, since it doesn't lead to the serious changes in the calculations
we shall take, as a lower limit, the Fermi scale $\,M_{F}=100\,GeV$.

\subsection{SU(N) model: running away from the finite theory}  

As an examples of the effect of the quantum conformal factor we
shall consider two toy models with two scalar fields: one which
possesses the one-loop finiteness without supersymmetry \cite{shya},
and another one which admits $N=2$ supersymmetry but is not finite.

Let us start with the generic model (\ref{action}).
The counterterms (\ref{divergency}) lead to the following expressions
for $\De\be_{p}$:
\beq
(4\pi)^2\,\De\beta_{f_1} =
12\alpha^2f_1 \,\zeta_1^2 \,+\, 4\alpha^4 \zeta_1^2
{(\zeta_1 - 1)}^2\,,\,\,\,\,\,\,\,\,\,\,\,\,\,\,\,\,\,\,\,\,\,\,\,\,
\,\,\,\,\,\,\,\,\,\,\,\,\,\,\,\,\,\,\,\,\,\,\,\,\,\,\,
\,\,\,\,\,\,\,\,\,\,\,\,\,\,\,\,\,\,\,\,\,\,\,\,\,\,\,\,
f_1(0) = f_1\,;
\label{nachalo}
\eeq
\beq
(4\pi)^2\,\De\beta_{f_2} =  12\alpha^2f_2 \,\zeta_1^2
\,,\,\,\,\,\,\,\,\,\,\,\,\,\,\,\,\,\,\,\,\,\,\,\,\,\,\,\,\,\,\,
\,\,\,\,\,\,\,\,\,\,\,\,\,\,\,\,\,\,\,\,\,\,\,\,\,\,\,\,\,\,\,\,\,\,\,\,
\,\,\,\,\,\,\,\,\,\,\,\,\,\,\,\,\,\,\,\,\,
\,\,\,\,\,\,\,\,\,\,\,\,\,\,\,\,\,\,\,\,\,\,\,\,\,\,\,
\,\,\,\,\,\,\,\,\,\,\,\,
f_2(0) = f_2\,;
\eeq
\beq
(4\pi)^2\,\De\beta_{f_3} =
 2\alpha^2\,f_3 \,(\zeta_1^2 \,+\, 4 \zeta_1\zeta_2 + \zeta_2^2)
 + 4\alpha^4 \zeta_1\zeta_2 \,(1 - \zeta_1 - \zeta_2 + \zeta_1 \zeta_2 )
\,,\,\,\,\,\,\,\,\,\,\,f_3(0) = f_3\,;
\eeq
\beq
(4\pi)^2\,\De\beta_{f_4} =
 2\alpha^2\,f_4 \,(\zeta_1^2 + 4 \zeta_1\zeta_2 + \zeta_2^2)
\,,\,\,\,\,\,\,\,\,\,\,\,\,\,\,\,\,\,\,\,\,\,\,\,\,\,
\,\,\,\,\,\,\,\,\,\,\,\,\,\,\,\,\,\,\,\,\,\,\,\,\,\,\,
\,\,\,\,\,\,\,\,\,\,\,\,\,\,\,\,\,\,\,\,\,\,\,\,\,\,\,\,\,\,\,\,\,\,\,\,\,\,
f_4(0) = f_4\,;
\eeq
\beq
(4\pi)^2\,\De\beta_{f_5} = 12\alpha^2\,f_5\, \zeta_2^2 +
4\alpha^4 \zeta_2^2 {(\zeta_2 - 1)}^2
\,,\,\,\,\,\,\,\,\,\,\,\,\,\,\,\,\,\,\,\,\,\,\,\,\,
\,\,\,\,\,\,\,\,\,\,\,\,\,\,\,\,\,\,\,\,\,\,\,\,\,\,\,
\,\,\,\,\,\,\,\,\,\,\,\,\,\,\,\,\,\,\,\,\,\,\,\,\,\,\,\,\,\,
f_5(0) = f_5\,\,\,
\eeq

and

\beq
(4\pi)^2\,\De\beta_{\zeta_j} =  2\al^2 \,\ze_j^2\, (\zeta_j - 1)
\,,\,\,\,\,\,\,\,\,\,\,\,\,\,\,\,\,\,\,\,\,\,\,\,\,
\,\,\,\,\,\,\,\,\,\,\,\,\,\,\,\,\,\,\,\,\,\,\,\,\,\,\,
\,\,\,\,\,\,\,\,\,\,\,\,\,\,\,\,\,\,\,\,\,\,\,\,\,\,\,\,\,\,
\,\,\,\,\,\,\,\,\,\,\,\,\,\,\,\,\,\,\,\,
\,\,\,\,\,\,\,\,\,\,\,\,\,\,
\zeta_i(0) = \zeta_i \,.
\label{xibetas}
\eeq

Let us now consider the asymptotic behavior  of the effective
couplings $f_i(t), \zeta_j(t)$. As it was already mentioned above,
all the  $\beta$-functions
(corresponding to gauge, Yukawa, scalar couplings and to the
nonminimal parameters $\xi$) vanish in the
conformal fixed point
\beq
h_k(t)=h_k^*  \,,\,\,\,\,\,\,\,\,\,\,\,\,\,\,
f_i(t)=f_i^*  \,,\,\,\,\,\,\,\,\,\,\,\,\,\,\,\zeta_j=0,
\label{fixed}
\eeq
where
$\,f_i^*,\,\,h_k^*\,$
are the values corresponding to the fixed point in flat space-time.
The corrections $\De\be$ indicate that there is also a
second "minimal" fixed point with the same (\ref{fixed}) solutions
for $h_k(t)$ and $f_i(t)$ but with $\zeta_j = 1$ (this corresponds
to the $\xi_j=0$ in (\ref{action}), that is why we call this fixed
point minimal). The behavior of the effective charges and effective
potential in the vicinity of the minimal fixed point has been studied
in \cite{foo} for the one-scalar model. Now we are interested in the
behavior of the effective charges close to the conformal fixed point.

As far as the renormalization group equations for $\,\,g(t),\,
h_k(t), \,f_i(t),\,
\zeta_j(t)\,\,$ are very cumbersome, it is reasonable to take
particular
models in which the study of these equations performs easier.
As a first example, let us take a particular model
of (\ref{action}) with $N=8$ and one scalar multiplet in the adjoint
representation. When the number of spinor
multiplets is $m=84$ the theory is one-loop finite
in the flat space-time
(see Ref.~\cite{shya} for a detailed  discussion of such a model
without taking the quantum effects of the conformal factor into account).
To perform the numerical analysis of the
renormalization group equations one has to impose the initial conditions
at far UV. We suppose that the running from the conformal fixed
point happens because of the two-loop contributions which violate the
$\xi=\frac16$ condition and thus switch on the interaction with the
conformal factor. Thus, to impose the initial conditions one has to
evaluate the two-loop effects. The two-loop contributions are, generally
speaking, proportional to $\frac{g^4}{(4\pi)^4}$ multiplied by the
combinatorial factor. Let us take the typical value of the parameter
$\zeta$ as $10^{-3}$, while the coupling constant is $g^2 = 0.1$.
Then the numerical analysis enables one to trace the behavior of the
effective couplings $\,\zeta(t),\,f_{1,2}(t)\,$ for

\noindent
$0>t=\ln ({\mu}/M_{Pl})>-39\,\,$
that corresponds to the running between
$\,\,\mu=M_{Pl}= 10^{19}\,GeV\,$ and $\,\mu=M_{F}= 10^{2}\,GeV$.

We have performed the numerical analysis of the full system of the
renormalization group equations for the effective
gauge, Yukawa, scalar couplings and $\,\ze(t)$ with taking into
account the corrections (\ref{nachalo}) -- (\ref{xibetas}).
As an initial point of the renormalization group flow
in far UV we choose the values (see \cite{shya} for the details)
$$
h_2=0\,;\,\,\,\,\,\,
f_1^* \approx 0.43621\,,\,\,\,\,\,\,f_2^* \approx 0.30277
$$
that provides the one-loop finiteness in the theory without back reaction
of  vacuum.
The results for the contribution to the couplings due to back-reaction,
that is the differences $Df_j$ and $D\ze_j$ between the couplings in the
with and without the back-reaction of vacuum,
are plotted in Fig.~\ref{oneSFDff}. It is easy to see that the
plots for the deviations look like a plane lines. The reason for this is
that the effect of the back reaction is very week and in such a ``small''
interval on the logarithmic scale the non-linear function looks like
a linear. Also the numerical values presented at the
Fig.~\ref{oneSFDff} show that the effect is very small. For instance,
the deviation of $f_{1,2}$ from the finite fixed point is about six
orders smaller than the value of the coupling itself in
this fixed point. One has to
notice that since the effect is almost linear, it is quite easy to
construct the same plots for other initial deviations $\,D\ze_1$.

The similar numerical analysis has been performed for the more
interesting $\,SU$(5) model (\ref{action}) - (\ref{rest}) with two
kinds of scalar fields and with
$$
m= 1\,, \,\, \,\, n=15\,;      \,\,\,\,\,\,\,\,\,\,\,\,\,\,\,\,
h_{1,2,3}^2(t) = h_{1,2,3}^*\,,\,\,\,\,\,\,\,\,\,\,\,\,\,\,\,\,
f_{1,...,5}^2(t) = f_{1,...,5}^*\,;
$$
where
$$
h^*_{1,2,3} = (1.421\,,1.681,\,2.361)\,;\,\,\,\,\,\,\,\,\,\,\,\,\,\,\,\,
\,\,\,\,
f^*_{1,2,3,4,5} = (0.659,\,1.293,\,0.324,\,1.677,\,1.039).
$$
The results of the numerical solution are presented
at the Fig.~\ref{twoSFDff} .The parameter $\,D\ze_1(0)$ here is taken
$0.01$ as in the previous example, and we kept the constraint
$\,D\ze_1 = D\ze_2\,$ for the sake of simplicity. Qualitatively
the results are the same as in the one-scalar case -- the effect exists
but it is tiny.


\pTwo{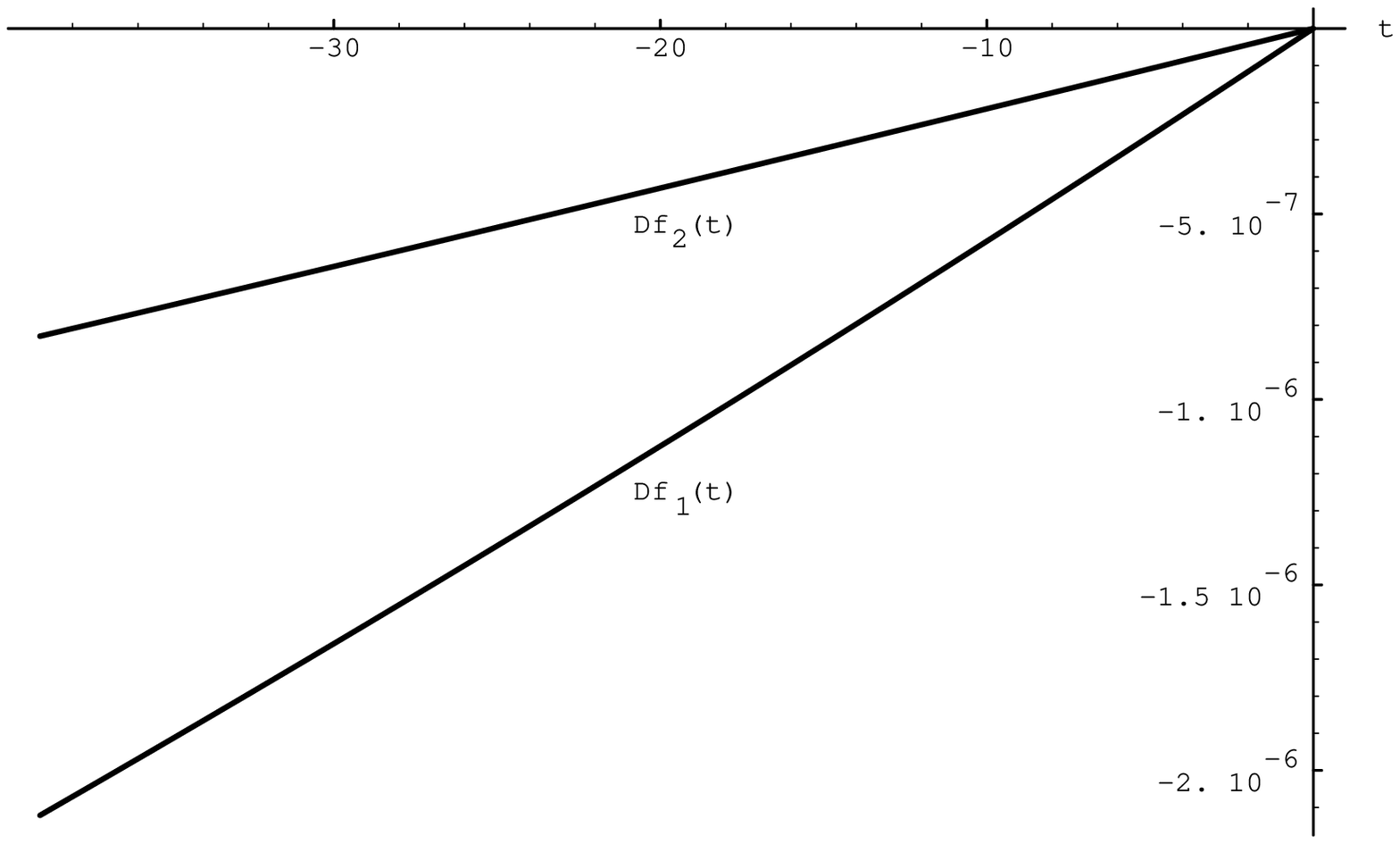}{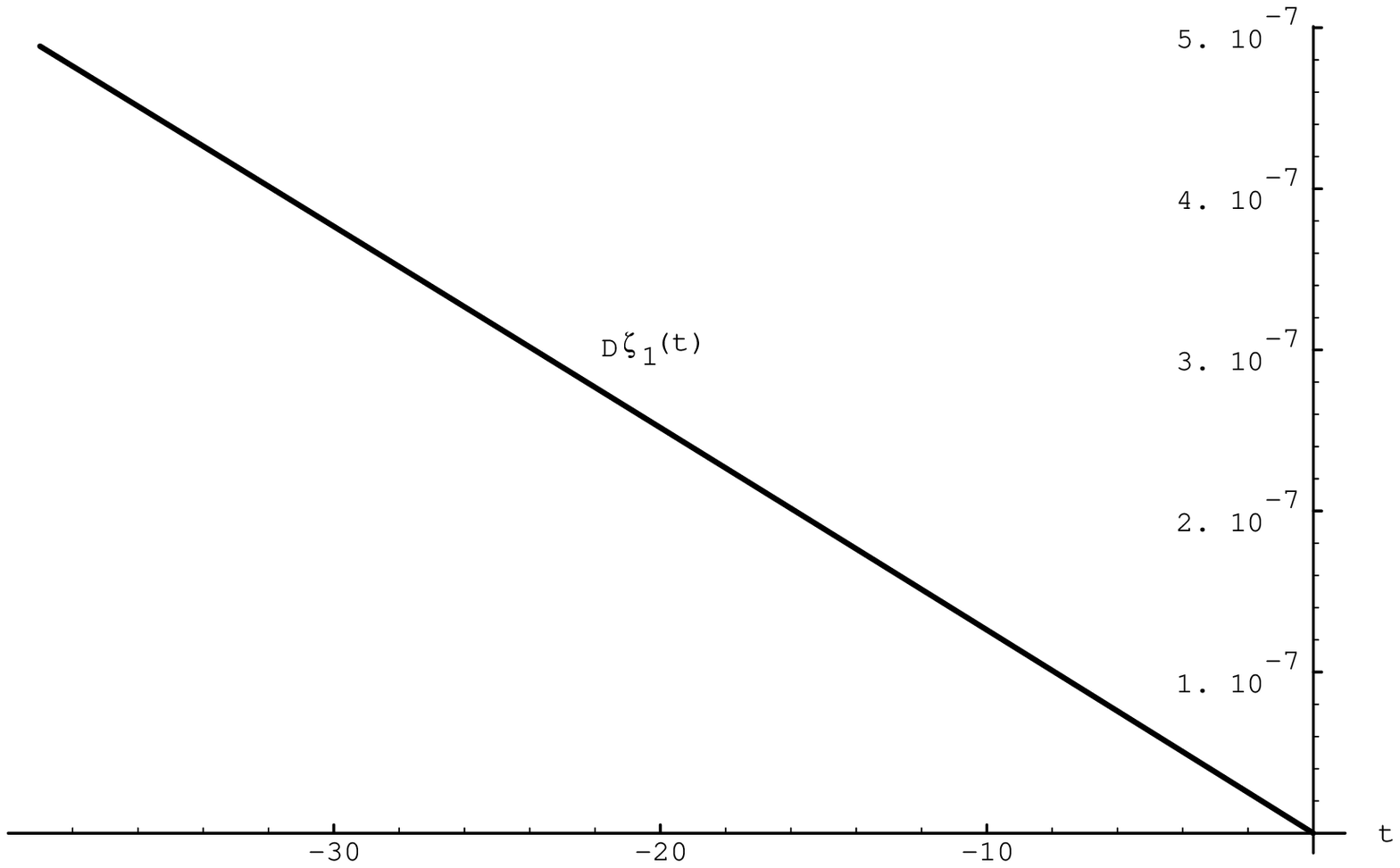}
     {Contribution to the running coupling constants due to back-reaction.
      $SU[8]$ model with one scalar field and 84 spinor multiplets.}


\pTwo{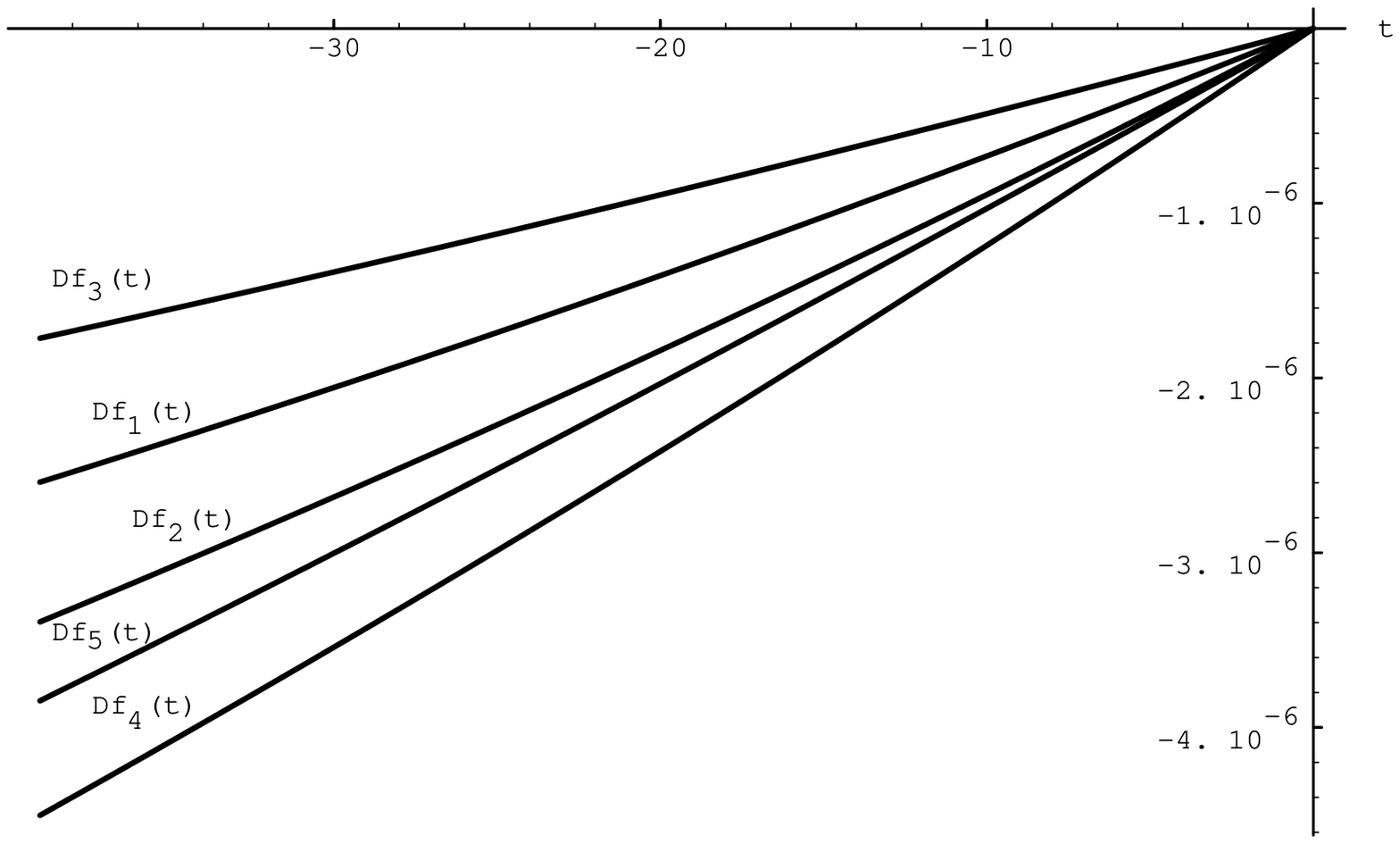}{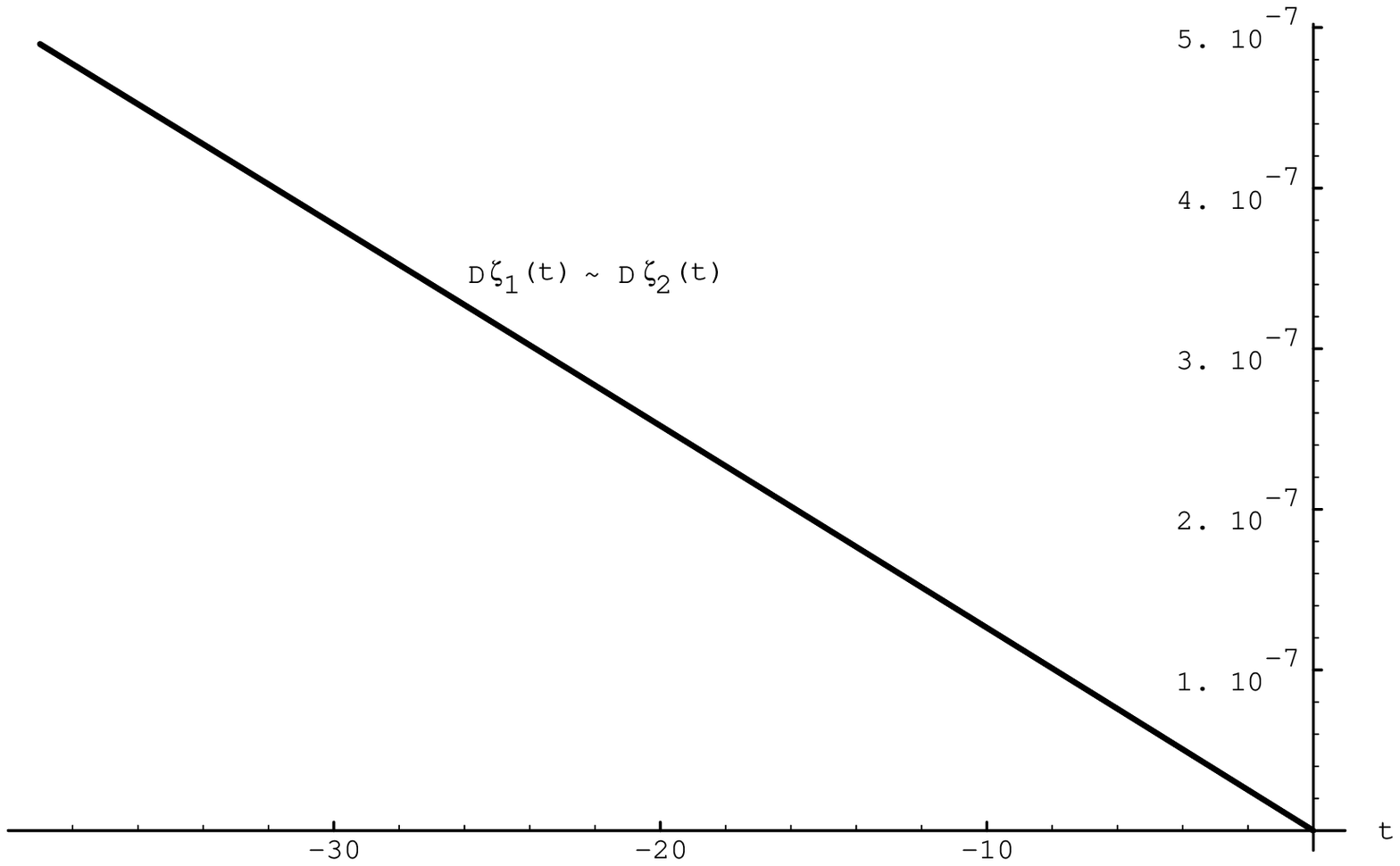}
     {Contribution to the running coupling constants due to back-reaction.
     $SU[5]$ model with two scalar fields and with the
     spinor content $m=1,\,\,n=15$.}

\subsection{Model with the rigid breaking of supersymmetry}

Let us now consider the $SU$(2) gauge model with an action
\cite{votyu,buod,book}:
$$
S\,=\,\int d^4\sqrt{-g}\,\left\{
-\frac14\,(G_{\mu\nu}^a)^2 +
\frac12\,g^{\mu\nu}\,\pa_\mu\ph\,\pa_\nu\ph +
\frac12\,g^{\mu\nu}\,\pa_\mu\La\,\pa_\nu\La
+\frac12\,\xi_1R\,\ph^2+\frac12\,\xi_2R\,\La^2 +
\right.
$$
$$
\left.
+ i\,{\bar \psi}_a\,\ga^\mu\,\na_\mu^{ab}\,\psi_b
- h_1{\bar \psi}_a \,F_{abc}\,\ph_b\,\psi_c
- i\,h_2{\bar \psi}_a \ga_5\,F_{abc}\,\La_b\,\psi_c +
\right.
$$
\beq
\left.
-\frac18\,f_1(\ph^2)^2 -\frac18\,f_2(\La^2)^2
-\frac14\,f_3\ph^2\La^2
-\frac12\,f_4\left[ \ph^2\La^2 - (\ph\La)^2 \right]
\,\right\}
\label{N=2}
\eeq
where $\ph$ and $\La$ are scalar and pseudoscalar fields in the
adjont representation of the $SU$(2) gauge group. Other notations are
obvious. In flat space, for the particular values of the couplings
\beq
f_1=f_2=f_3=0,\,\,\,\,\,\,f_4=h_1^2=h_2^2=g^2
\label{ini}
\eeq
the model (\ref{N=2}) possesses the $N=2$ supersymmetry.
For other values of the couplings one meets the renormalizable
theory which has also second non-supersymmetric renormalization
group fixed point \cite{votyu}. In curved space time this model is
also renormalizable (if only the nonminimal terms and the action
of vacuum are introduced), and admits the asymptotic conformal
invariance for (\ref{ini}) and $\xi_1+\xi_2 = 1/6$.

Taking into account our previous results we arrive at the
renormalization group equations for the above model with the back
reaction of vacuum.
The gauge coupling behaves as \cite{votyu}
$$
g^2(t) = g^2\,\left(1+\frac{8g^2t}{(4\pi)^2}\right)^{-1}
$$
and supersymmetry fixes Yukawa constants to be $h_1^2=h_2^2=g^2$. In
flat
space-time those relations hold under renormalization \cite{votyu}.
The same happens when we put our theory in curved space-time and
take the back reaction of vacuum into account, because the
quantum conformal factor contributes to the $\be$-functions of the
scalar couplings and nonminimal parameters.
We are interested in the renormalization group flow from the
supersymmetric and conformal invariant UV fixed point,
that is why we have to impose the constraints (\ref{ini})
on the initial data. As far as the back reaction doesn't concern
the behavior of the Yukawa couplings, the relations (\ref{ini})
for the Yukawa couplings
hold for all scales and they can be indeed used in the RG equations
directly, while $f_{1,2,3,4}$ should be regarded as arbitrary
quantities. After we use the
relations between gauge and Yukawa coupling constants,
the RG equations for $f_{1,2,3,4}(t)$ become
$$
(4\pi)^2\,\frac{df_1}{dt}\, =\,
11f_1^2+3f_3^2+8f_3f_4+8f_4^2-8f_1g^2-8g^4
+ 12\al^2f_1\ze_1^2 + 4\al^4\ze_1^2(\ze_1 - 1)^2\,,\,\,\,\,\,\,\,\,\,\,\,
$$
$$
(4\pi)^2\,\frac{df_2}{dt}\, =\,
11f_2^2+3f_3^2+8f_3f_4+8f_4^2-8f_2g^2-8g^4
+ 12\al^2f_2\ze_2^2 + 4\al^4\ze_2^2(\ze_2 - 1)^2\,,\,\,\,\,\,\,\,\,\,\,\,
$$
$$
(4\pi)^2\,\frac{df_3}{dt}\, =\,
4f_3^2+(5f_3+4f_4)(f_1+f_2) +8f_4^2 - 8f_3g^2 - 8g^4
+ 2\al^2f_3(\ze_1^2 + 4 \ze_1\ze_2 + \ze_2^2) +
$$$$
 +\,\, 8\,\al^4\,\ze_1 \ze_2\,(1 - \ze_1 - \ze_2 + \ze_1\ze_2)\,,
\,\,\,\,\,\,\,\,\,\,\,\,\,\,\,\,\,\,\,\,\,\,\,\,\,\,\,\,\,\,\,\,\,\,\,
\,\,\,\,\,\,\,\,\,\,\,\,\,\,\,\,\,\,\,\,\,\,\,\,\,\,\,\,\,\,\,\,\,\,\,
\,\,\,\,\,\,\,\,\,\,\,\,\,\,\,\,\,\,\,
$$
\beq
(4\pi)^2\,\frac{df_4}{dt} \,=\,
6f_4^2 - 8f_4g^2 - 6g^4 + 8f_3f_4 +2 f_4(f_1+f_2)
+ 2\al^2f_4(\ze_1^2 + 4 \ze_1\ze_2 + \ze_2^2)\,,\,\,\,
\label{equa}
\eeq
while for $\xi_{1,2}(t)$ we meet the equations:
$$
(4\pi)^2\,\frac{d\ze_1}{dt} \,=\,
\ze_1\,(5f_1 - 4g^2) + \ze_2 (3f_3+4f_4)
+ 2\al^2\ze_1^2(\ze_1 - 1)\,,\,\,\,\,\,\,\,\,\,\,\,\,
$$
\beq
(4\pi)^2\,\frac{d\ze_2}{dt} =
\ze_2\,(5f_2 - 4g^2) + \ze_1 (3f_3+4f_4)
+ 2\al^2\ze_2^2(\ze_2 - 1)\,.
\label{eqSUSY}
\eeq

In order to study the renormalization group behaviour of the
effective couplings $\,f(t),\,\,\xi(t)\,$ we have to choose the
initial value of $g^2$ and the initial variation
$\,D\ze_1=-D\ze_2\neq 0$.
As in the previous case we take in the far UV $\,g^2 = 0.1\,$ and
$\,\ze_1 = 0.001\,$.
The results of the numerical analysis, concerning the
contributions due to back-reaction, are plotted in
Figs.~\ref{su001Dff} for the case $\ze_1=0.001$.


\pTwo{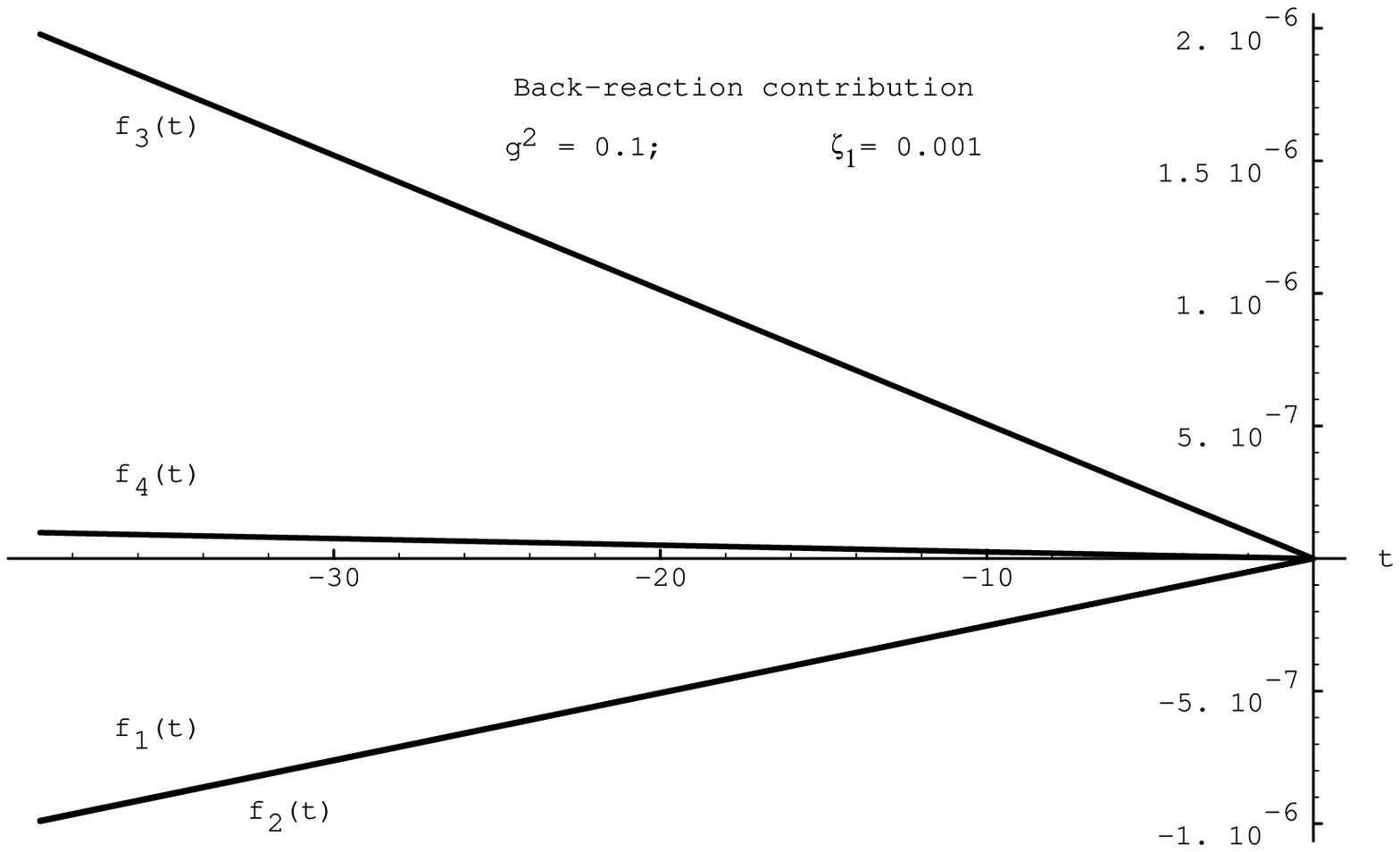}{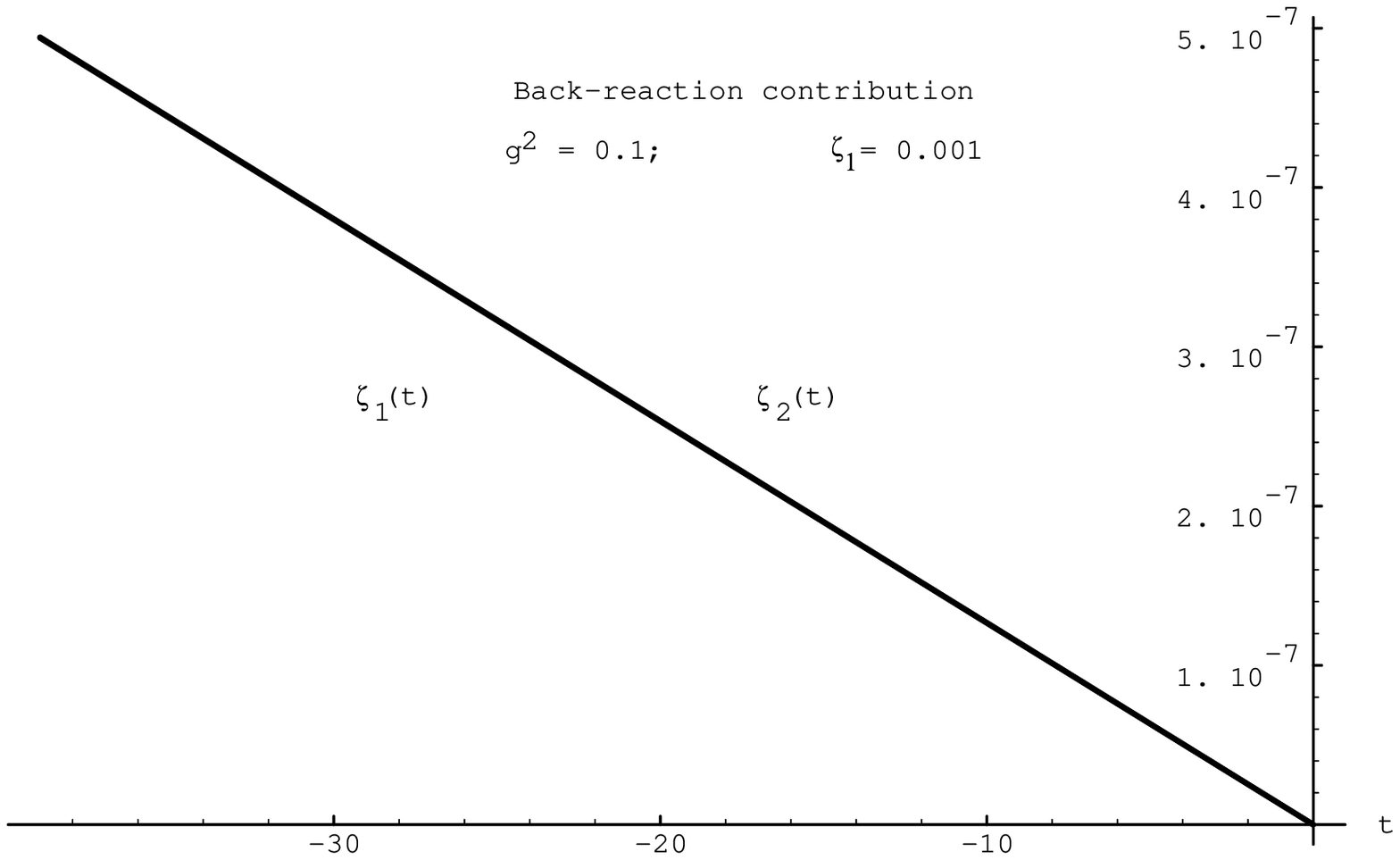}
     {Contribution of back-reaction to the running couplings
      $f_j(t)$ and $\ze_j(t)$.
      First example with $\ze_1(0)=-\ze_2(0)=0.001$.}


One can see that in this case the effect of quantum conformal factor
is essentially the same as in the previous case of the $SU$(8) and
$SU$(5) models. Because of those effects one meets the
nonzero $f_{1,2,3}$ couplings with the first two having the values
about $10^{-8}$ at the $M_X$ unification scale and about $10^{-6}$
at the $M_F$ Fermi scale, while they remain identically zero without
the back reaction of the conformal factor.

In order to give some speculation on the strongly coupled theories
we also investigated the case with initial values of $g^2 = 0.5$
and $\ze=0.1$. The results are plotted at Fig. ~\ref{su1Dff}.

\pTwo{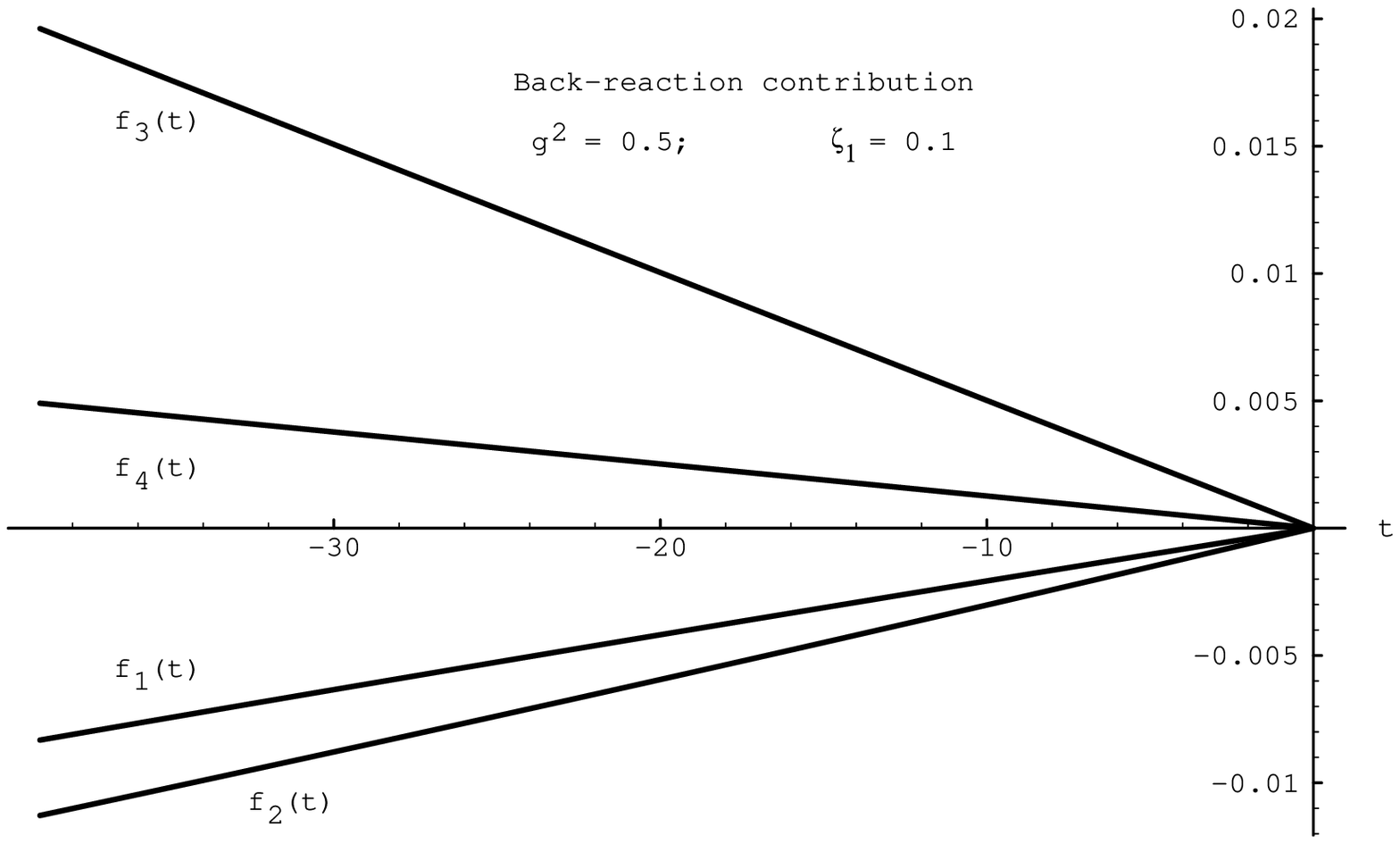}{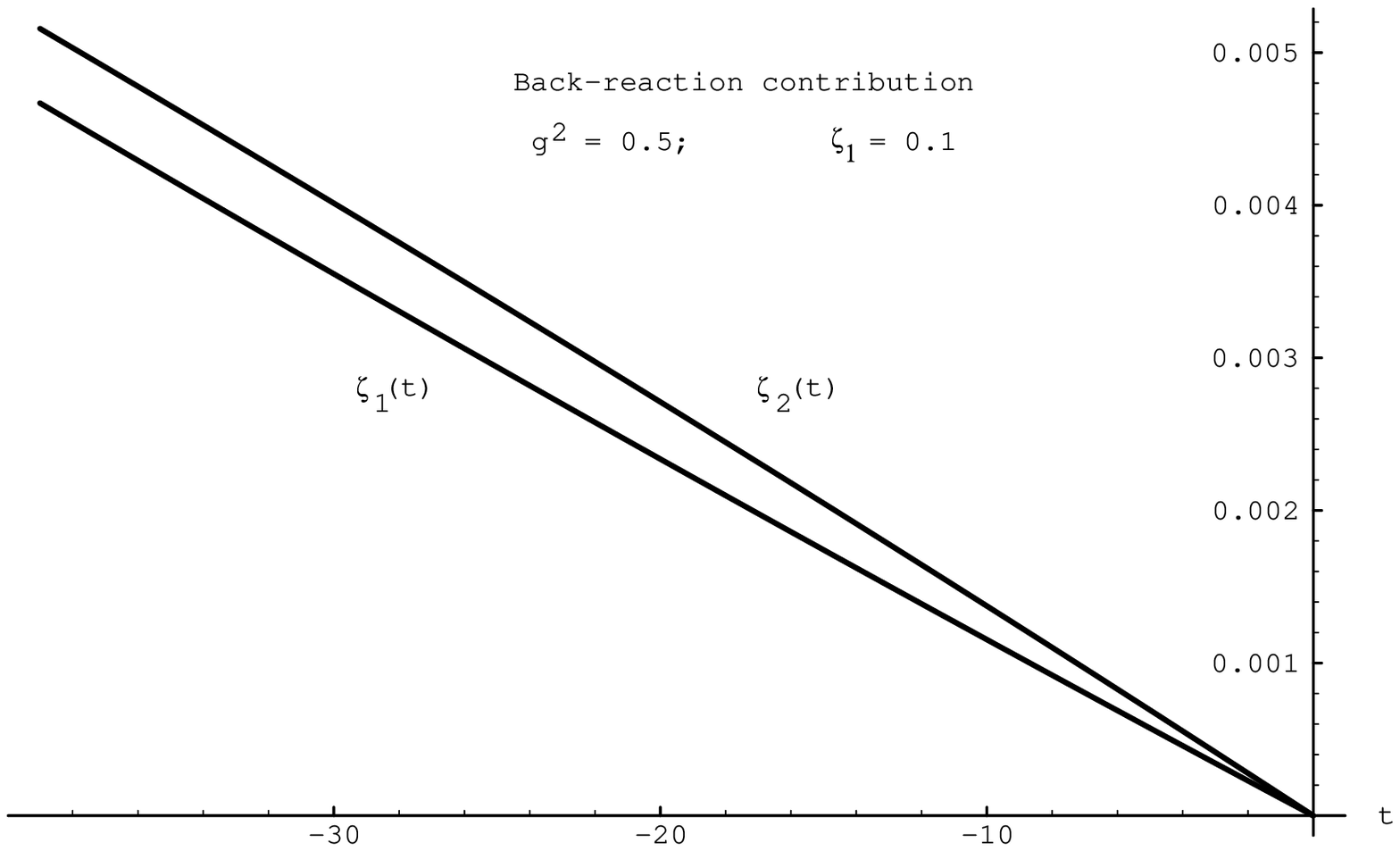}
     {Contribution of back-reaction to the running couplings
      $f_j(t)$ and $\ze_j(t)$.
      $g^2=0.5$ and $\ze_1(0)=-\ze_2(0)=0.1$.}

\pTwo{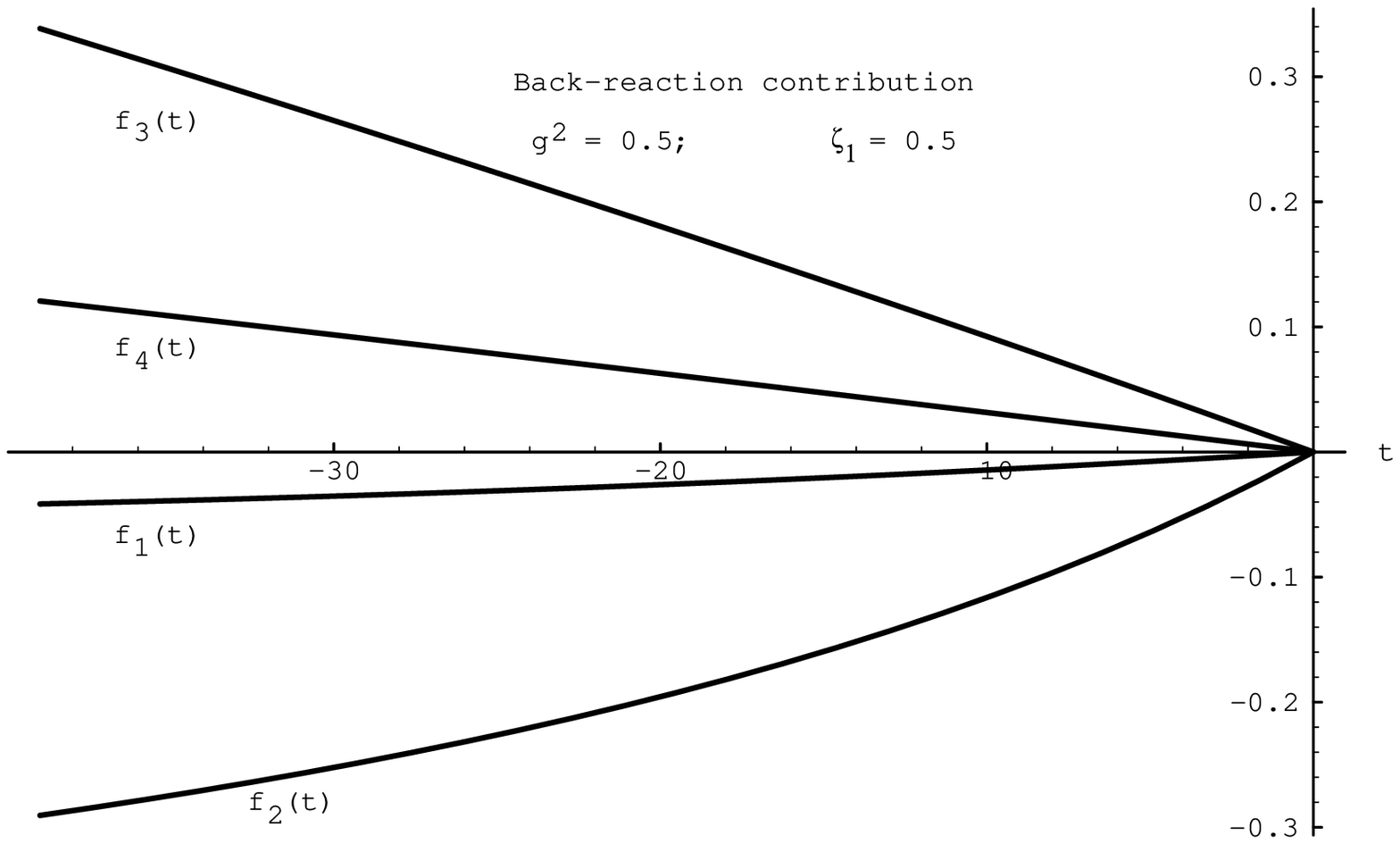}{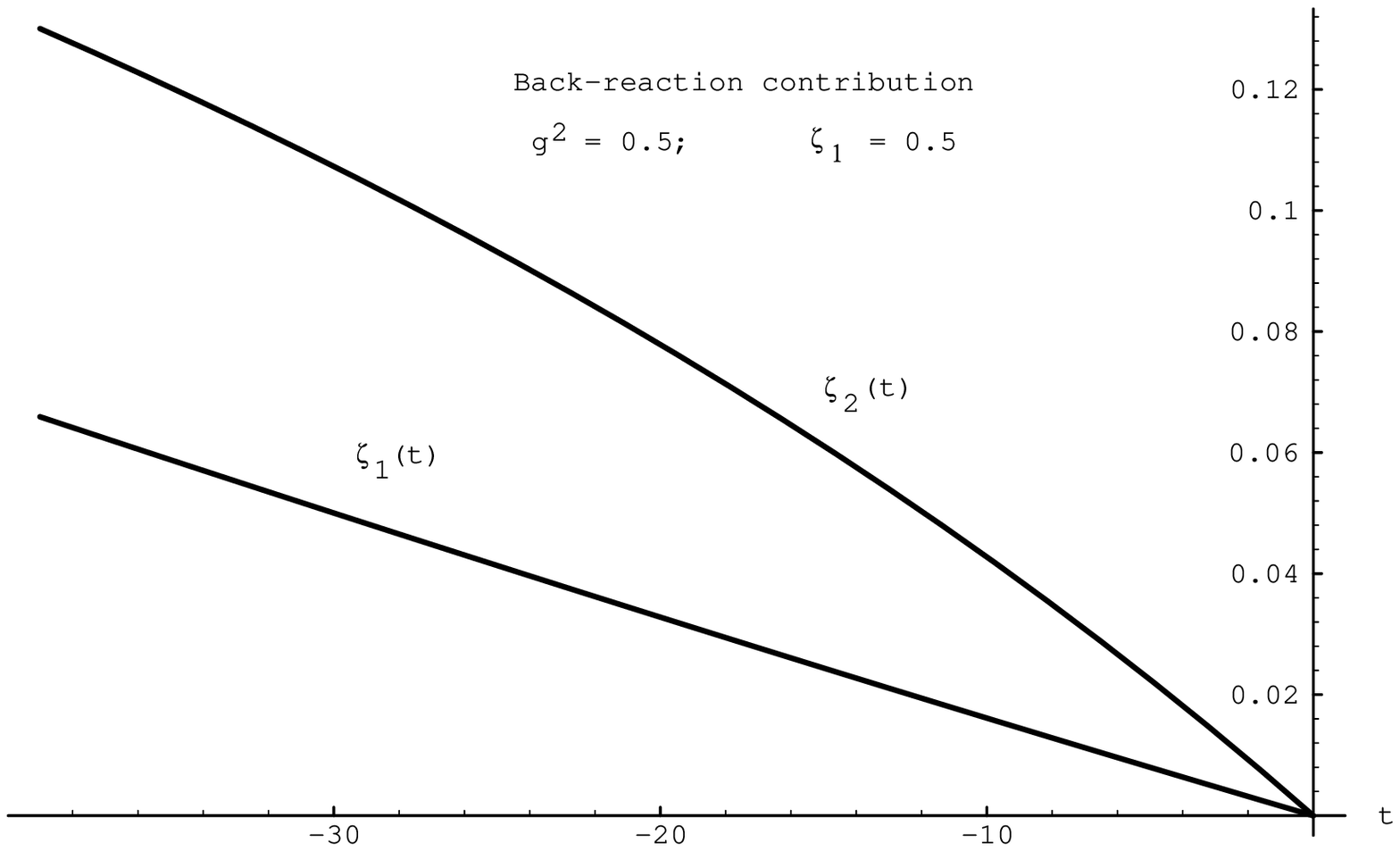}
     {Contribution of back-reaction to the running couplings
      $f_j(t)$ and $\ze_j(t)$.
      $g^2=0.5$ and $\ze_1(0)=-\ze_2(0)=0.5$.
This example was included only to illustrate the nonrealistic
order of magnitude for $g^2$ and initial deviations $D\zeta_i$
for which the nonlinear behavior shows up.}

\section{Conclusion}

The arguments were presented that in four-dimensional quantum field
theory in curved space-time the local conformal symmetry can not be
exact, and that it can be only approximate symmetry which serves as
the initial point for the renormalization group equations in the far
UV limit. Starting from the conformal invariant theory one meets
the trace anomaly and consequent propagation of the conformal factor
of the metric.
We have considered the interaction between the quantized conformal
factor and the matter fields in a theory with approximate
conformal symmetry, in a region close to the scale of asymptotic
freedom and asymptotic conformal invariance.

The contributions
 of the conformal factor to the $\beta$-functions of
the scalar coupling constants have been calculated and it was shown that
these contributions drive the theory out from the conformal point and
also modify a values of the scalar coupling
constants. If the starting model
in far UV possesses supersymmetry and (or) finiteness, the interaction
with the conformal factor leads to the violation of those
properties at lower energies.
For the initial violations about $D\ze_i=0.001$
the numerical values of the deviations of the scalar couplings from
the symmetric state range from $10^{-8}$ to $10^{-6}$ at the
Fermi scale and about one order less at the Unification scale.
Indeed the effect is very weak and therefore all the dependencies are
very close to be linear. In particular the above deviations may be
considered as proportional to the squares of the
initial violations $(D\ze_i)^{2}$ of the
conformal invariance. At higher loops
the supersymmetry breaking in the scalar sector should violate
supersymmetry in the Yukawa and gauge interactions. As a result
the relations between the masses of the particles at low energies
may differ from the ones which are expected in supersymmetric
GUT's and the supersymmetry is not seen at low energies.
However, since the numerical effect of such a violation will be
extremely small, it is very difficult to see how one can observe it.
One can say that if the local conformal invariance exists as a
high energy symmetry, it holds as a very good approximation at
lower energies.

\newpage
\section{Acknowledgments} I.Sh. thanks Departamento de Fisica, Universidade
Federal de Juiz de Fora for warm hospitality. The work of I.Sh. has been
supported in part by CNPq (Brazil) and by the Russian 
Foundation for Basic Research under the project No.96-02-16017.

\end{document}